%% file: Template.tex
\documentclass{article}
\usepackage{spconf,amsmath,graphicx,hyperref}
\usepackage{orcidlink}
\usepackage{amssymb,amsfonts}
\usepackage{adjustbox, booktabs}
\usepackage{tikz}
\usetikzlibrary{arrows.meta,positioning,fit}
\usepackage[nolist, nohyperlinks]{acronym}
\usepackage[table]{xcolor}
\usepackage{comment}

\title{Domain-Incremental Learning for Multi-channel Replay Speech Detection}
\name{Michael~Neri~\orcidlink{0000-0002-6212-9139}}
\address{\textit{Faculty of Information Technology and Communication Sciences, Tampere University, Tampere, Finland}\\
         \textit{michael.neri@tuni.fi}\\
}

\begin{acronym}
\acro{AE}{autoencoder}
\acro{ASV}{automatic speaker verification}
\acro{BCE}{binary cross-entropy}
\acro{CM}{countermeasure}
\acro{CL}{continual learning}
\acro{CNN}{convolutional neural network}
\acro{COTS}{commercial-off-the-shelf}
\acro{CQCC}{constant-Q cepstral coefficient}
\acro{CRNN}{convolutional recurrent neural network}
\acro{DNN}{deep neural network}
\acro{DF}{deepfake}
\acro{DIL}{domain-incremental learning}
\acro{EER}{equal error rate}
\acro{ELU}{exponential linear unit}
\acro{FCN}{fully convolutional network}
\acro{FFT}{fast Fourier transform}
\acro{GMM}{Gaussian mixture model}
\acro{GRU}{gated recurrent unit}
\acro{ID}{identifier}
\acro{IoT}{Internet of things}
\acro{LA}{logical access}
\acro{LSTM}{long short-term memory}
\acro{MLP}{multilayer perceptron}
\acro{MVDR}{minimum variance distortionless response}
\acro{MWF}{multichannel Wiener filtering}
\acro{ReMASC}{Realistic Replay Attack Microphone Array Speech}
\acro{RNN}{recurrent neural network}
\acro{ROC}{receiving operating characteristic}
\acro{TSB}{task-specific beamformer}
\acro{TTS}{text-to-speech}
\acro{PA}{Physical access}
\acro{PFA}{probability of false alarm}
\acro{SOTA}{state-of-the-art}
\acro{SRP-PHAT}{steered response power with phase transform}
\acro{STFT}{short-time Fourier transform}
\acro{SVD}{singular value decomposition}
\acro{VA}{voice assistant}
\acro{VC}{voice conversion}
\acro{EWC}{Elastic weight Consolidation}
\acro{GPM}{Gradient Projection Memory}
\acro{AE}[AE]{Average Accuracy}
\acro{AIE}[AIE]{Average Incremental Accuracy}
\acro{fm}[FM]{Forgetting Measure}
\acro{bwt}[BWT]{Backward Transfer}
\acro{im}[IM]{Intransigence Measure}
\acro{fwt}[FWT]{Forward Transfer}
  
\end{acronym}

\begin{document}
\ninept

\newcommand{\mich}[1]{{\textbf{\textcolor{red_cool}{#1}}}}
\newcommand{\meanci}[2]{#1\,{\scriptsize$\pm$#2}}
\definecolor{red_cool}{rgb}{0.5, 0.0, 0.0}
\definecolor{CLImprove}{HTML}{0072B2}  
\definecolor{CLWorsen}{HTML}{E69F00}  

\maketitle
\begin{abstract}
Replay attacks are the most accessible threat to voice-controlled systems, and the acoustic cues that expose them are strongly modulated by the environment in which the attack is mounted. A detector deployed in the field therefore has to absorb new acoustic conditions over time, ideally without revisiting past recordings, since retaining speech indefinitely is both expensive and legally constrained. We frame this as \ac{DIL} over acoustic environments and present the first continual learning benchmark for multi-channel replay speech detection, evaluating a state-of-the-art beamformer-based detector over all $24$ environment orderings of the ReMASC corpus with five seeds. Sequential fine-tuning forgets severely, raising the error rate on previously learned environments by $18.8$ points. Elastic weight consolidation (EWC) halves forgetting but loses plasticity, gradient projection memory (GPM) is statistically indistinguishable from naive fine-tuning, and the proposed task-specific beamformer (TSB) that keeps one spatial front-end per environment significantly improves final and incremental accuracy. We further show that the last environment of the sequence dominates final performance. Code, results, and analysis are available at \textit{\href{https://github.com/michaelneri/replay-speech-continual}{https://github.com/michaelneri/replay-speech-continual}}.
\end{abstract}
\section{Introduction}
\label{sec:intro}
 
Voice-controlled systems are now a standard interface to smart speakers, smartphones, and voice-based banking, where \ac{ASV} separates a legitimate user from an impostor. Among the attacks that defeat it, replay is by far the most accessible~\cite{Wu_SpeechComm_2015, Kinnunen_Interspeech_2017}. In fact, an adversary only needs to record a victim's utterance and play it back to the target device, with no technical expertise and no access to synthesis or conversion models. A replayed utterance goes through one acquisition-reproduction cycle more than a genuine one, so traces are left at the spoofing stage as well. The attacker's microphone and the playback loudspeaker add their own frequency response and non-linear distortion, and the reverberation of the room where the utterance was captured is already imprinted in the signal reaching the target device. Detectors look for these traces, and recent ones exploit the multiple microphones already present in commercial devices to extract spatial cues that a single channel cannot provide~\cite{Neri_EUSIPCO_2025, Neri_OJSP_2025, Meng_USENIX_2022}.

The cues separating the two classes are spatial. A talker is a compact source at a distance and orientation typical of a person addressing the device, whereas a loudspeaker sits closer to the array, radiates with a different directivity, and emits an already reverberated signal, so that inter-channel time and level differences, coherence, and direct-to-reverberant ratio differ between the two. These cues are modulated by the enclosure itself, hence they manifest differently in an office, in a corridor, or inside a moving car, and detectors degrade when the recording conditions change with respect to training~\cite{Neri_EUSIPCO_2025}. A deployed device keeps encountering new acoustic conditions, and retraining from scratch on the union of all data is impractical. Additionally, storing past recordings for later reuse is also undesirable and impractical, as speech is personal and it is classified as biometric data~\cite{gdpr2016}.

Continual learning~\cite{VANDEVEN2025153, French_TICS_1999} addresses exactly this scenario, in which a model is updated on a stream of data increments. Updating a replay detector as new acoustic environments are encountered one after the other is a domain-incremental problem, the scenario in which the label space is fixed and only the input distribution changes. Unlike generalisation to unseen environments, the detector is expected to learn each environment it is deployed in while retaining the previous ones~\cite{todisco2019asvspoof, Neri_EUVIP_2022}. An exemplar-free regularisation approach was introduced in~\cite{Ma_Interspeech_2021} to adapt a detector to new attacks without revisiting past data, and subsequent work refined the weight-modification strategy to better balance stability and plasticity~\cite{Zhang_ICML_2023, Zhang_AAAI_2024}, with further contributions using feature distillation and class balancing~\cite{Wani_ICPR_2025}. These works, however, are single-channel and attack-incremental: each increment introduces new spoofing algorithms while the acoustic conditions remain comparatively fixed. Physical-access replay poses a different problem, since the discriminative cue is modulated by the recording environment itself, which prior work has addressed with domain adaptation between environments~\cite{Wang_Interspeech_2020}, but not incrementally.

\input{figures/tsb}

The contributions of this work are as follows: (i) to the best of our knowledge, this is the first work to frame \textbf{multi-channel physical-access replay detection} as a \textbf{domain-incremental problem over acoustic environments}; (ii) we define the \ac{TSB}, an architecture-based procedure that instantiates one learnable beamformer per environment on top of a shared classification \ac{CRNN} and requires no environment label at inference; (iii) we benchmark a state-of-the-art multi-channel detector together with \ac{TSB} and two standard continual learning approaches over all environment orderings and five seeds, and show that the order of the environments affects the outcome more than the choice of algorithm.

\section{Method}
\label{sec:materials}
This section first introduces the problem of the \ac{DIL} for replay speech detection. Then, it describes the components of the study: the multi-channel replay speech detector we adopt and the continual learning strategies we compare, including the proposed one.

\subsection{DIL for Replay Speech Detection}
Replay speech detection is the binary classification of a multi-channel recording as genuine, i.e., uttered by a talker in front of the array, or replayed, i.e., a previously captured utterance reproduced by a loudspeaker. In this work a detector is updated on a stream of $K$ acoustic environments, observed one at a time and never jointly: since only the input distribution changes across increments, the setting is domain-incremental. At step $k$ the model is initialised from the previous set of parameters and optimised on the new environment alone, without revisiting past utterances and without access to the environment identity at inference.

\subsection{Replay Speech Detector}
\label{ssec:model}
In our experiments we employ the state-of-the-art replay speech detector M-ALRAD~\cite{Neri_OJSP_2025}, a \ac{CRNN} that jointly processes $N$ complex \acp{STFT} $\{X_{{\mathrm{STFT}}_{n,T,F}}, n = 1, \ldots, N\}$ of a multi-channel recording, with $T$ and $F$ time and frequency bins respectively, to produce a single-channel beamformed spectrogram. Each recording is resampled to $16$~kHz, cropped to its first second, zero-padding shorter utterances, and peak-normalised before computing the \acp{STFT} with a $1024$-point window and $50\%$ overlap, which gives $T=32$ and $F=513$. First, a \ac{CNN} $f_{BM}:\mathbb{C}^{N \times T \times F} \rightarrow \mathbb{C}^{T \times F}$ predicts the beamforming weights $\mathbf{W} \in \mathbb{C}^{N \times T \times F}$ through a Conv2D-BatchNorm-ELU-Conv2D stack, where real and imaginary parts are concatenated along the channel dimension. The beamformed spectrogram is then computed as 
$\hat{X}_{{\mathrm{STFT}_{T, F}}} = \sum_{n} X_{{\mathrm{STFT}}_{n,T,F}} \cdot
\mathbf{W}_{n,T,F}$. $\hat{X}_{\mathrm{STFT}}$ is classified by a \ac{CRNN}, denoted with $g_\phi$ with $\phi$ the set of learnable parameters,  previously used for speaker distance estimation~\cite{Neri_TASLP_2024, Neri_WASPAA_2023}. Its magnitude, together with the sine and cosine of its phase, forms a $T \times F \times 3$ tensor that is passed through three convolutional layers with $1 \times 3$ filters and batch normalization, each followed by parallel max and average pooling. Two bi-directional \ac{GRU} layers with $128$ neurons refine the feature maps, and the final hidden state is mapped to the binary prediction $\hat{y} \in \mathbb{R}^2$ by a fully connected layer with softmax activation function, where $0$ denotes the genuine class and $1$ the replay one. Training minimizes the label-weighted binary cross-entropy loss, together with the orthogonality and sparsity losses on $\mathbf{W}$ using the same hyperparameters as in~\cite{Neri_OJSP_2025}.

\subsection{Spatial Continual Learning Approach}
We propose task-specific beamformer (\ac{TSB}), an architecture-based continual learning strategy that decouples M-ALRAD~\cite{Neri_OJSP_2025} into an environment-specific spatial front-end and an environment-agnostic classifier, shown in Figure~\ref{fig:tsb}. This design follows the observation that the multi-channel cues exploited by the beamformer are strongly environment-dependent (array geometry, reverberation, noise field).

Reusing the beamformer $f_{BM}:\mathbb{C}^{N\times T\times F}\!\to\! \mathbb{C}^{T\times F}$, we instantiate $K$ independent heads $\{f^{(k)}_{BM}\}_{k=1}^{K}$, one per domain/environment, each with parameters $\theta^{(k)}$ and producing a beamformed spectrogram $\hat{X}^{(k)} = f^{(k)}_{BM}(X_{\mathrm{STFT}})$. All heads share a single \ac{CRNN} classifier $g_\phi$, mapping a beamformed spectrogram to the class logits $\mathbf{z}=g_\phi(\hat{X}^{(k)})\in\mathbb{R}^{2}$. When the model is adapted to the $k$-th domain, only its head $\theta^{(k)}$ and the shared \ac{CRNN} classifier $\phi$ are optimised, while all earlier heads $\{\theta^{(i)}\}_{i<k}$ are frozen: 
\begin{equation}
    \theta^{(k)},\phi = \arg\min_{\theta^{(k)}, \phi} \mathcal{L}\left(g_\phi\left(\hat{X}^{(k)}\right),\,y\right),
\end{equation}
where $\mathcal{L}$ is the binary weighted cross-entropy with the orthogonality and sparsity regularisers, as in~\cite{Neri_OJSP_2025}. Freezing past heads makes the spatial filters of earlier environments immutable, structurally preventing forgetting at the beamforming stage, while the shared classifier stays fully plastic so the discriminative cues keep improving. At inference no domain label is required: the input is processed by all trained heads and their logits are averaged before the softmax,
\begin{equation}
    \hat{y} = \mathrm{softmax}\left(
    \frac{1}{K}\sum_{k=0}^{K-1} g_\phi\left(\hat{\mathbf{X}}^{(k)}\right)
    \right),
\end{equation}
where at an intermediate domain only the heads trained so far take part in the average. The overhead over the baseline is one head per domain, growing linearly with $K$; each $f^{(k)}_{BM}$ adds only ${\approx}9.4$k parameters, a small fraction of the shared classifier.

\subsection{Continual Learning Baselines}
In this work, we focus on regularization-based, optimization-based, and architecture-based continual learning approaches. We discard replay- and representation-based~\cite{Wang_TPAMI_2024} techniques because (i) speech is personal and biometric data under the GDPR~\cite{gdpr2016}, hence retaining utterances of past environments for the whole life of the model conflicts with data minimisation, storage limitation and the right to erasure; and (ii) representation-based methods are built on large-scale pre-trained or self-supervised encoders, which for multi-channel spatial audio do not exist in the literature. We adapt two continual learning strategies to the multi-channel replay speech detection task: (i) \ac{EWC}~\cite{kirkpatrick2017ewc} (regularization-based) penalises movement away from the previous solution through $\tfrac{\lambda}{2}\sum_i F_i(\theta_i-\theta^{\star}_{k-1,i})^2$, where $F$ is the diagonal empirical Fisher accumulated over one pass on the environment just learned; we keep a single anchor, so the carried state stays at $2|\theta|$ scalars, and set $\lambda=5\,000$. (ii) \ac{GPM}~\cite{saha2021gpm} (optimization-based) constrains the direction of the update instead: after each step the per-batch parameter gradients of every layer are stacked and a \ac{SVD} retains the leading vectors covering $\varepsilon_{\mathrm{th}}=0.97$ of the singular-value energy; the resulting bases are extended online and later gradients are projected onto their orthogonal complement, $\mathbf{g}\leftarrow\mathbf{g}-MM^{\top}\mathbf{g}$. Both EWC and GPM hyperparameters are reported in Table~\ref{tab:cl_hparams}.

\begin{table}[t]
\caption{Configuration of the \ac{DIL} benchmark: setting shared by all methods (top) and method-specific hyper-parameters (bottom).}
\label{tab:cl_hparams}
\centering
\begin{adjustbox}{max width=\columnwidth}
\begin{tabular}{ll}
\toprule
\multicolumn{2}{l}{\textit{Shared setting}}\\
\midrule
Input & 4-channel, $1$\,s, $16$\,kHz (from $44.1$\,kHz)\\
STFT & $N_{\mathrm{FFT}}=1024$, hop $=512$\\
Optimiser & Adam, $\eta=10^{-4}$, weight decay $10^{-4}$\\
Scheduler & cosine annealing, $T_{\max}=100$, $\eta_{\min}=10^{-5}$\\
Batch / epochs per env. & $8$ / $50$, gradient clipping $1.0$\\
Model selection & lowest EER on the held-out partition\\
Sequences & $4!=24$ orderings $\times$ $5$ seeds $=120$\\
\midrule
\multicolumn{2}{l}{\textit{Method-specific}}\\
\midrule
EWC & $\lambda=5\,000$; diagonal empirical Fisher, single anchor\\
GPM & $\varepsilon_{\mathrm{th}}=0.97$, residual tol.\ $0.1$; parameter-gradient bases\\
TSB & $K=4$ heads, $9\,416$ params.\ each ($+11.3\%$); logit ensemble\\
\bottomrule
\end{tabular}
\end{adjustbox}
\end{table}

\section{Continual Learning Benchmark Design}
We define here how the approaches of Section~\ref{sec:materials} are evaluated. Detection performance at each step is measured with the \ac{EER}, while the metrics below summarise an entire incremental sequence in terms of accuracy, forgetting, and transfer.


\subsection{Continual Learning Metrics}
Let $K$ denote the number of sequentially-learned environments ($K=4$ in our experiments), indexed $0,\dots,K-1$, and let $\mathrm{E}[k,j]$ denote the \ac{EER} on environment $j$'s test set after the model has been trained sequentially on environments $0,\dots,k$ (only $j \le k$ is meaningful). It is worth noting that for the computation of the metrics the indices $j,k$ denote the position of an environment within a given ordering, not a fixed environment identifier. Following~\cite{Wang_TPAMI_2024}, we define the following metrics:

\textbf{AE — Average Error.} The mean \ac{EER} across all $K$ environments after the full training sequence has completed
\begin{equation}
\mathrm{AE} = \frac{1}{K}\sum_{j=0}^{K-1} \mathrm{E}[K{-}1,j].
\end{equation}
AE summarises overall error rate at deployment time, after the model has seen every environment. 

\textbf{AIE — Average Incremental Error.} Since AE only looks at the final step, AIE rewards models that perform well \emph{throughout} training, not only at the end. Let $\overline{\mathrm{AE}}_k$ be the running average at intermediate step $k$. Then, we compute the mean over all steps 
\begin{equation}
\overline{\mathrm{AE}}_k = \frac{1}{k+1}\sum_{j=0}^{k} \mathrm{E}[k,j],
\qquad
\mathrm{AIE} = \frac{1}{K}\sum_{k=0}^{K-1} \overline{\mathrm{AE}}_k .
\end{equation}

\textbf{FM — Forgetting Measure.} For each environment $j$ learned before the final step, FM compares its \emph{best-ever} \ac{EER} (over all steps where it could be evaluated, up to but excluding the final step) against its \ac{EER} \emph{at} the final step:
\begin{equation}
\mathrm{FM} = \frac{1}{K-1}\sum_{j=0}^{K-2}
\Big( \mathrm{E}[K\!-\!1,j] - \min_{j\leq i\leq K-2} \mathrm{E}[i,j] \Big).
\end{equation}
A positive FM means the \ac{EER} on environment $j$ rose above its best previously-achieved value by the end of training, i.e., the model forgot what it had learned about $j$. 

\textbf{BWT — Backward Transfer.} BWT measures how much the \ac{EER} on each previously-learned environment $j$ changed between the step at which $j$ was first learned and the end of the full sequence:
\begin{equation}
\mathrm{BWT}
= \frac{1}{K-1}\sum_{j=0}^{K-2}\big(\mathrm{E}[j,j] - \mathrm{E}[K{-}1,j]\big).
\end{equation}
A negative BWT means the final EER on $j$ is higher than it was when $j$ was just learned, i.e., forgetting occurred. 

\textbf{IM — Intransigence Measure.} IM requires a jointly-trained reference model: $\mathrm{E}_{\mathrm{ref}}[k]$ is the \ac{EER}, evaluated on environment $k$, of a model trained on environments $0,\dots,k$ \emph{simultaneously}. IM is the gap between the sequential (continual) model and the reference model at the final step:
\begin{equation}
\mathrm{IM} = \mathrm{E}[K{-}1,K{-}1] - \mathrm{E}_{\mathrm{ref}}[K{-}1].
\end{equation}
A positive IM means the continually-trained model could not reach the performance a jointly-trained model achieves on the same data. 

\begin{table*}[th!]
\caption{Continual learning benchmark on ReMASC (D2, $16$ kHz), pooled over $24$ orderings $\times$ 5 runs ($n=120$). $\downarrow$/$\uparrow$ = lower/higher is better; best per column in bold. AE and AIE in \%, others in percentage points. Values are mean $\pm$ 95\% CI half-width ($t$-distribution). Significance is assessed against the naive fine-tuning baseline with Wilcoxon signed-rank tests paired by (ordering, run) ($^{*}p<0.05$, $^{**}p<0.01$, $^{***}p<0.001$); shading marks a significant ($p_W<0.05$) \colorbox{CLImprove!20}{\strut improvement} / \colorbox{CLWorsen!20}{\strut degradation}.}
\label{tab:cl_global}
\centering
\begin{adjustbox}{max width=0.9\textwidth}
\begin{tabular}{lcccccc}
\toprule
Algorithm & AE $\downarrow$ & AIE $\downarrow$ & FM $\downarrow$ & BWT $\uparrow$ & IM $\downarrow$ & FWT $\uparrow$ \\
\midrule
Naive fine-tuning (baseline)
  & \meanci{24.09}{1.15}
  & \meanci{19.56}{0.76}
  & \meanci{18.79}{1.58}
  & \meanci{-18.66}{1.59}
  & \meanci{-4.34}{1.09}
  & \meanci{3.42}{0.62} \\
EWC~\cite{kirkpatrick2017ewc}
  & \meanci{23.61}{1.13}
  & \meanci{19.92}{1.11}
  & \cellcolor{CLImprove!20}\meanci{\textbf{10.18}$^{***}$}{1.45}
  & \cellcolor{CLImprove!20}\meanci{\textbf{-9.63}$^{***}$}{1.45}
  & \cellcolor{CLWorsen!20}\meanci{1.84$^{***}$}{1.47}
  & \cellcolor{CLWorsen!20}\meanci{-5.12$^{***}$}{1.08} \\
GPM~\cite{saha2021gpm}
  & \meanci{24.15}{1.13}
  & \meanci{19.69}{0.77}
  & \meanci{18.45}{1.58}
  & \meanci{-18.36}{1.59}
  & \meanci{-4.17}{1.08}
  & \meanci{3.42}{0.74} \\
TSB (ours)
  & \cellcolor{CLImprove!20}\meanci{\textbf{23.06}$^{*}$}{1.14}
  & \cellcolor{CLImprove!20}\meanci{\textbf{18.75}$^{***}$}{0.82}
  & \meanci{17.77}{1.56}
  & \meanci{-17.64}{1.54}
  & \cellcolor{CLImprove!20}\meanci{\textbf{-6.69}$^{**}$}{1.70}
  & \meanci{\textbf{4.04}}{0.65} \\
\bottomrule
\end{tabular}
\end{adjustbox}
\end{table*}

\textbf{FWT — Forward Transfer.} FWT requires single-task reference models: $\mathrm{E}_{\mathrm{sref}}[j]$ is the \ac{EER} of a model trained \emph{only} on environment $j$, with no exposure to any other environment. For every environment except the first, FWT compares this single-task baseline against the \ac{EER} achieved by the continual model at the step where $j$ was first introduced
\begin{equation}
\mathrm{FWT} = \frac{1}{K-1}\sum_{j=1}^{K-1}
\big(\mathrm{E}_{\mathrm{sref}}[j] - \mathrm{E}[j,j]\big).
\end{equation}
A positive FWT means the model, having already learned earlier environments, performs \emph{better} on a new environment $j$ than a model trained on $j$ alone, i.e., prior environments provided useful forward transfer. 

\section{Experimental Results}
\label{sec:results}

We first introduce the corpus which has been used to carry out the experiments. Then, we report the outcome of the benchmark over the $24$ environment orderings and five seeds, first comparing the four strategies on the continual learning metrics and then isolating the effect of the position that each environment occupies in the sequence.

\subsection{Dataset}
Experiments are conducted on \ac{ReMASC}~\cite{Gong_Interspeech_2019}, the only publicly available corpus providing synchronised multi-channel recordings of replay attacks, collected with four microphone arrays across four acoustic environments: an outdoor scenario (Env-A), two enclosed spaces (Env-B and Env-C), and a moving vehicle (Env-D).

Among the four arrays we retain only $\mathrm{D}2$, a linear array of four omnidirectional microphones sampled at $44.1$~kHz, and leave the remaining ones to future work. Fixing the array makes the acoustic environment the only axis along which the domain changes: the arrays differ in the number of microphones and in geometry, which would alter the input dimensionality of the beamformer and confound the spatial mismatch with the environment shift. $\mathrm{D}2$ provides $10{,}664$ utterances ($2{,}452$ genuine and $8{,}212$ replay), distributed over Env-A to Env-D as $1{,}942$, $3{,}614$, $2{,}423$, and $2{,}685$ samples respectively, and we adopt the speaker-disjoint partition released with the dataset, with $40$ speakers for training and $11$ for testing.

\subsection{Results}
Each algorithm is compared against the naive fine-tuning baseline with paired tests matched by (ordering, run), exploiting the fact that all algorithms are evaluated on the identical set of $4! \times 5 = 120$ environment-ordering and run combinations. Pairing removes the substantial variance attributable to ordering difficulty, isolating the effect of the algorithm itself. Following~\cite{demsar2006}, we adopt the Wilcoxon signed-rank test~\cite{wilcoxon1945} as the decisive criterion, as \ac{EER} distributions are bounded and right-skewed. Table~\ref{tab:cl_global} shows that sequential fine-tuning forgets severely. Specifically, the \ac{EER} of the baseline on the environments it has already learned rises by $18.79$ points, and the sequence closes at $24.09\%$ average \ac{EER}. \ac{TSB} is the only method that significantly improves the error-oriented metrics, reducing AE by $1.02\pm0.92$ points and AIE by $0.81\pm0.42$, while FM and BWT remain statistically unchanged: per-environment beamformer heads lower the \ac{EER} by adapting to the acoustics of each environment rather than by preventing forgetting. \ac{TSB} also attains the best intransigence ($-2.35\pm1.30$ against the baseline), i.e., sequential training over-specialises to the environment seen last and surpasses the joint reference on it.

\ac{EWC} exhibits the expected stability-plasticity trade-off. The Fisher-based penalty anchors the weights to the previous optimum and nearly halves forgetting ($10.18$ against $18.79$ points of FM, $p_W<0.001$), but the same rigidity prevents the model from reaching the joint-training optimum ($+6.18$ points of IM) and produces negative forward transfer ($-8.54$ points of FWT), showing that the penalties of previous environments interfere with the adaptation to new ones. \ac{GPM}, instead, is statistically indistinguishable from the baseline on all six metrics: projecting gradients onto the null space of the subspaces of previous environments brings no measurable benefit when all environments share the same binary objective and their gradient directions largely overlap.

\subsection{Analysis on the environment order}

\begin{figure}[th!]
    \centering
    \includegraphics[width=0.87\columnwidth]{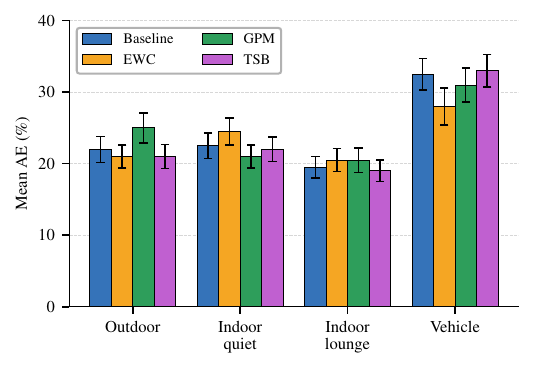}
    \caption{Effect of the last environment of the sequence on the mean AE (95\% CI, $t$-distribution), averaged over all orderings ending with that environment ($6\times5=30$ observations per bar). The four environments are the outdoor scenario (traffic and wind), the quiet enclosed space, the indoor lounge (background music and TV), and the moving vehicle.}\vspace{-5mm}
    \label{fig:ordering}
\end{figure}

Figure~\ref{fig:ordering} shows the mean AE as a function of the last environment in the \ac{DIL} sequence, averaged across all orderings ending with that environment and all five runs. The results reveal that the final environment in the sequence is the dominant factor driving overall performance, while the choice of first environment has negligible influence. Ending the sequence on the vehicle environment (moving car, non-stationary engine and road noise) consistently yields the highest \ac{EER} across all four methods, with mean AE between $28\%$ and $33\%$, compared to $19-21\%$ when the indoor lounge environment (stationary background music and TV) is last. This pattern is primarily a data-driven effect: the vehicle environment is intrinsically the most acoustically challenging condition, and sequential fine-tuning on it last causes the model to over-adapt to its distribution at the expense of previously learned environments. 

Notably, EWC partially mitigates this effect, achieving a mean AE of approximately $28\%$ on the vehicle-last condition compared to $32$--$33\%$ for the other methods. This is consistent with EWC's Fisher-based regularisation, which prevents excessive weight drift when adapting to the final environment, inadvertently preserving representations learned from earlier and easier conditions. Environment ordering is therefore a practically relevant deployment variable, and EWC offers partial robustness to a difficult final condition at the cost of plasticity.

\section{Conclusion}
We presented the first continual learning benchmark for multi-channel replay speech detection, framing the acoustic environments of \ac{ReMASC} as a domain-incremental sequence and evaluating four exemplar-free strategies over all $24$ orderings with five seeds. None of them removes catastrophic forgetting. \ac{EWC} halves the forgetting measure but pays for it with negative forward transfer, \ac{GPM} is indistinguishable from naive fine-tuning on all six metrics, and the proposed \ac{TSB} improves final and incremental accuracy without reducing forgetting, indicating that its gain comes from spatial specialisation rather than from added stability. The environment that closes the sequence drives the final \ac{EER} far more than the algorithm does. Future work will address increments over microphone arrays with different geometries and numbers of channels, and combine spatial specialisation with an explicit stability mechanism.

\bibliographystyle{IEEEbib}
\bibliography{refs}

\end{document}

%% file: figures/tsb.tex
\definecolor{TSBact}{HTML}{0072B2}   
\definecolor{TSBfrz}{HTML}{8A8A8A}

\begin{figure}[t]
\centering
\begin{adjustbox}{max width=0.9\columnwidth}
\begin{tikzpicture}[
  font=\scriptsize,
  >={Stealth[length=1.5mm]},
  dim/.style={font=\scriptsize,inner sep=1pt},
  head/.style={draw,rounded corners=1.5pt,minimum width=14mm,minimum height=5.4mm,
               inner sep=1pt},
  frz/.style={head,draw=TSBfrz,fill=TSBfrz!15,text=black!60},
  act/.style={head,draw=TSBact,fill=TSBact!18,thick},
  io/.style={draw,rounded corners=1.5pt,minimum width=13mm,minimum height=7mm,
             inner sep=1.5pt,align=center},
  bb/.style={draw=TSBact,thick,rounded corners=1.5pt,minimum width=16mm,
             minimum height=35.6mm,align=center,fill=TSBact!8},
  sum/.style={draw,circle,inner sep=0pt,minimum size=7mm}
]

\node[io] (x) at (0,0)
  {$\mathbf{X}_{\mathrm{STFT}}$\\[-1pt]{\tiny $\mathbb{C}^{N\times T\times F}$}};

\node[frz] (b1) at (2.3, 1.29) {$f^{(1)}_{\mathrm{BM}}$};
\node[frz] (b2) at (2.3, 0.43) {$f^{(2)}_{\mathrm{BM}}$};
\node      (bd) at (2.3,-0.43) {$\vdots$};
\node[act] (b3) at (2.3,-1.29) {$f^{(k)}_{\mathrm{BM}}$};
\node[draw=black!45,dashed,rounded corners=2pt,inner sep=2.2mm,
      fit=(b1)(b3)] (bank) {};
\node[font=\tiny,above=0.4mm of bank] {task-specific beamformer heads};
\node[dim,below=1mm of bank] {$\hat{X}^{(i)}\in\mathbb{C}^{T\times F}$};

\node[bb] (h) at (4.4,0) {shared\\ CRNN \\ classifier\\[2pt]$g_{\phi}$};
\node[dim,below=1mm of h] {$z\in\mathbb{R}^{2}$};

\node[sum] (s) at (6.35,0) {$\frac{1}{K}\!\sum$};
\node[io]  (y) at (8.4,0)  {$\hat{y}$\\[-1pt]{\tiny genuine / replay}};

\draw (x.east) -- (1.15,0);
\draw (1.15,-1.29) -- (1.15,1.29);
\foreach \i/\yy in {1/1.29, 2/0.43}
  \draw[->,dashed,TSBfrz] (1.15,\yy) -- (b\i.west);
\draw[->,TSBact,thick] (1.15,-1.29) -- (b3.west);

\foreach \i/\yy in {1/1.29, 2/0.43}
  \draw[->,dashed,TSBfrz] (b\i.east) -- (3.6,\yy);
\draw[->,TSBact,thick] (b3.east) -- (3.6,-1.29);

\foreach \yy in {1.29, 0.43, -1.29}
  \draw[->,black!55] (5.2,\yy) -- (5.65,\yy);
\draw[black!55] (5.65,-1.29) -- (5.65,1.29);
\draw[->] (5.65,0) -- (s.west);
\draw[->] (s.east) -- node[above,font=\tiny]{softmax} (y.west);

\node[frz,minimum width=4mm,minimum height=3mm,inner sep=0pt] (l1) at (1.6,-2.75) {};
\node[right=1mm of l1,font=\tiny] (t1) {$\{\theta^{(i)}\}_{i<k}$ frozen};
\node[act,minimum width=4mm,minimum height=3mm,inner sep=0pt,
      right=4mm of t1] (l2) {};
\node[right=1mm of l2,font=\tiny] {$\theta^{(k)},\phi$ optimised};

\end{tikzpicture}
\end{adjustbox}
\caption{Task-specific beamformer (TSB). When the model is adapted to the $k$-th domain, only the head $\theta^{(k)}$ and the shared classifier $\phi$ are optimised (solid path), while all earlier heads $\{\theta^{(i)}\}_{i<k}$ are frozen (dashed paths). At inference the logits of the heads trained so far are averaged before the softmax, so no domain label is required.}
\label{fig:tsb}\vspace{-5mm}
\end{figure}